\documentclass[conference]{IEEEtran}
\IEEEoverridecommandlockouts

\usepackage{siunitx}
\usepackage{authblk}

\usepackage{booktabs, makecell}

\usepackage[colorlinks=true, urlcolor=magenta]{hyperref}

\usepackage{subcaption}

\usepackage{cite}
\usepackage{amsmath,amssymb,amsfonts}
\usepackage{algorithmic}
\usepackage{graphicx}
\usepackage{textcomp}
\usepackage{xcolor}
\usepackage{url}

\def\BibTeX{{\rm B\kern-.05em{\sc i\kern-.025em b}\kern-.08em
    T\kern-.1667em\lower.7ex\hbox{E}\kern-.125emX}}
\begin{document}

\title{Malware Detection in Docker Containers: \\ An Image is Worth a Thousand Logs \\

\thanks{*This project has received funding from the European Union’s Horizon Europe research and innovation programme under grant agreement No 101070455 (DYNABIC). Disclaimer: Funded by the European Union. Views and opinions expressed are however those of the author(s) only and do not necessarily reflect those of the European Union or European Commission. Neither the European Union nor the European Commission can be held responsible for them.
}

%
%
\thanks{
\IEEEauthorrefmark{1} A. Nousias, E. Katsaros, E. Syrmos and P. Radoglou-Grammatikis are with K3Y Ltd, Sofia, Bulgaria -- email: 
\texttt{\{anousias, ekatsaros, esyrmos, pradoglou\}@k3y.bg}
}

\thanks{
\IEEEauthorrefmark{3} T. Lagkas is with the Department of Computer Science, International Hellenic University, Kavala, Greece -- email: 
\texttt{tlagkas@cs.ihu.gr}
}

\thanks{
\IEEEauthorrefmark{4} V. Argyriou is with the Department of Networks and Digital Media, Kingston University, London, U.K. -- email: 
\texttt{vasileios.argyriou@kingston.ac.uk}
}

\thanks{
\IEEEauthorrefmark{5} I. Moscholios is with the Department Informatics \& Telecommunications, University of Peloponnese, Tripolis, Greece -- email: \texttt{idm@uop.gr}
}

\thanks{
\IEEEauthorrefmark{6} S. Goudos is with the Physics Department, Aristotle University of Thessaloniki, Thessaloniki, Greece -- email: \texttt{sgoudo@physics.auth.gr}
}
\thanks{
\IEEEauthorrefmark{7} E. Markakis is with the Department of Electrical and Computer Engineering, Hellenic Mediterranean University, Heraklion, Greece - email: \texttt{emarkakis@hmu.gr}
}
\thanks{
\IEEEauthorrefmark{2} P. Radoglou-Grammatikis and P. Sarigiannidis are with the Department of Electrical and Computer Engineering, University of Western Macedonia, Kozani, Greece - email: 
\texttt{\{pradoglou, psarigiannidis\}@uowm.gr}\
}


}

\author{
    Akis Nousias\IEEEauthorrefmark{1}, Efklidis Katsaros\IEEEauthorrefmark{1}, Evangelos Syrmos\IEEEauthorrefmark{1}, 
    Panagiotis Radoglou-Grammatikis\IEEEauthorrefmark{1}\IEEEauthorrefmark{2},
    Thomas Lagkas\IEEEauthorrefmark{3},\\
    Vasileios Argyriou\IEEEauthorrefmark{4},
    Ioannis Moscholios\IEEEauthorrefmark{5}, 
    Evangelos Markakis\IEEEauthorrefmark{7},
    Sotirios Goudos\IEEEauthorrefmark{6} and
    Panagiotis Sarigiannidis\IEEEauthorrefmark{2}
}

\maketitle

\begin{abstract}
Malware detection is increasingly challenged by evolving techniques like obfuscation and polymorphism, limiting the effectiveness of traditional methods. Meanwhile, the widespread adoption of software containers has introduced new security challenges, including the growing threat of malicious software injection, where a container, once compromised, can serve as entry point for further cyberattacks. In this work, we address these security issues by introducing a method to identify compromised containers through machine learning analysis of their file systems. We cast the entire software containers into large RGB images via their tarball representations, and propose to use established Convolutional Neural Network architectures on a streaming, patch-based manner. To support our experiments, we release the COSOCO dataset--the first of its kind--containing 3364 large-scale RGB images of benign and compromised software containers at~\href{https://huggingface.co/datasets/k3ylabs/cosoco-image-dataset}{https://huggingface.co/datasets/k3ylabs/cosoco-image-dataset}. Our method detects more malware and achieves higher F1 and Recall scores than all individual and ensembles of VirusTotal engines, demonstrating its effectiveness and setting a new standard for identifying malware-compromised software containers.
\end{abstract}

\begin{IEEEkeywords}
cybersecurity, malware detection, deep learning, docker containers
\end{IEEEkeywords}

\section{Introduction}

In the field of cybersecurity, the growing complexity and sophistication of malware present a significant and evolving challenge to detection mechanisms. Traditionally, malware detection has heavily relied on signature-based methods, where input files are compared against databases of known malware. While this approach has been effective, it is becoming increasingly inadequate as modern malware employ adaptive techniques such as obfuscation and polymorphism.
Signature-based methods are particularly vulnerable to zero-day attacks from novel malware variants that have not yet been observed in the wild, as they can only detect threats that have already been cataloged. This limitation highlights the need for adaptive detection mechanisms~\cite{scott2017signature}. In response, research is increasingly shifting towards machine learning-based methods~\cite{yu2021securing}, which can better capture emerging and novel threats. 

Moreover, as new technologies and practices emerge, cybercriminals continually adapt, finding new methods for delivering and executing malicious software. In recent years, software containers have gained widespread adoption in software engineering, primarily due to their ability to abstract system-specific dependencies and their scalability. Each container represents a standardized, self-contained unit of software and can support a wide range of functionalities and applications, from operating systems, and databases to machine learning models, among others. Their adoption, while beneficial, presents significant security challenges. One growing threat is the injection of malicious software into containers, which once compromised, can serve as entry points for further attacks.

In this paper, we investigate the use of machine learning-based techniques for identifying malware-compromised containers in an end-to-end manner. Previous research has primarily focused on using machine learning to classify executable files as either malware or benign or to analyze memory dumps and system calls. In contrast, we treat the entire container's file system as input and explore whether machine learning can detect changes inflicted by malware. The key question we aim to answer is: \textit{"Is this system compromised?"}. By analyzing the container's file system as a whole, we attempt to capture fine-grained characteristics that may be overlooked when monitoring separate files, system resources or calls.

We propose a novel approach that extends the use of Convolutional Neural Networks (CNNs) to dockerized software containers by converting their file systems into large RGB images on which the CNN operates. Malware typically operate locally affecting only a small subset of bytes within a system leaving a vastly larger number of bytes mainly unaltered. Thus, in the context of a system as an image, the pixels altered by malware are only a small fraction of the total. This scenario is similar to the task of classifying very large images with small objects, where the focus is to detect regions of interest within a much larger and more complex background that is non-informative of the label~\cite{kong2021efficientclassificationlargeimages}.
Additionally, we introduce and publicly release a novel dataset of benign and malware-compromised container images, along with an extensible data generation pipeline and evaluation protocol. This pipeline supports various malware types, processor architectures, and Operating Systems (OS), to enable further research across varied environments. Finally, we evaluate multiple CNN architectures to assess their effectiveness on this new input type. Our contributions are summarized as follows:
\begin{itemize}
    \item \textbf{Task Formulation:} We define the novel task of identifying malware-compromised dockerized software containers with machine learning-based methods and propose a streaming, patch-based CNN approach motivated by Multiple Instance Learning (MIL)~\cite{Carbonneau_2018}.
    \item \textbf{Compromised Software Containers (COSOCO) Image Dataset:} We introduce a novel dataset containing 3,364 large-scale RGB image representations of benign and compromised dockerized software containers.
    \item \textbf{Data Generation Pipeline:} We present a fully customizable and scalable data generation pipeline for creating images of benign and compromised software containers across various OS and CPU architectures.
\end{itemize}

\section{Background \& Related Work}


\subsection{Malware Detection Overview}
\label{Malware Detection Overview}

Malware detection techniques~\cite{aslan2020comprehensive} rely on well-established methods such as signature-based detection, behavioral monitoring, heuristic analysis, and sandboxing. Signature-based detection~\cite{sathyanarayan2008signature} remains the most widely used method in modern antivirus software, where the system scans files and programs for patterns that match known malware signatures. This approach is fast for detecting known malware, but struggles with novel or polymorphic malware. Heuristic-based detection analyzes the static features of a file or program before execution. It examines patterns, or commands within the code to identify malicious behavior. Heuristics look for suspicious characteristics, like obfuscation, uncommon instructions, or anomalies in the code itself, without running the program. It uses rules to identify potentially malicious behavior, offering some defense against unknown malware but often leading to false positives.  
Behavioral monitoring goes beyond static analysis by observing the runtime behavior of applications and processes, detecting actions commonly associated with malware, such as unauthorized network communication or file modifications such as registry changes. While effective, this technique may miss dormant malware that becomes active later. Sandboxing provides a safer environment by executing suspicious files in isolated environments to analyze their behavior without risking system integrity. Although this method is effective, it is resource-intensive and can be circumvented by malware that recognizes such environments.

In contrast to previous methods that typically investigate individual files for malicious content - before, during, or after runtime - our work focuses on entire docker containers. Therefore, it reformulates the question \textit{"Is this file malicious?"} to \textit{"Is this system compromised?"}. It can be viewed as a mixture of behavioral monitoring, heuristic analysis while bearing sandboxing characteristics. It is similar to heuristic-based detection in that it analyzes the static features of a file, yet this file is the entire software container. It is similar to behavioral monitoring in that it examines file system modifications such as registry changes, yet it does so by inspecting the whole tarball file instead of the runtime activity of a prespecified, ``under-investigation" program. Finally, our work touches upon sandbox environments via the data generation pipeline which utilizes isolation technologies to restrict the malware activity locally, yet integrates the containers with other technologies to ``fool" malware into activation.

\begin{figure*}[!htbp]
    \centering
    \includegraphics[width=\linewidth]{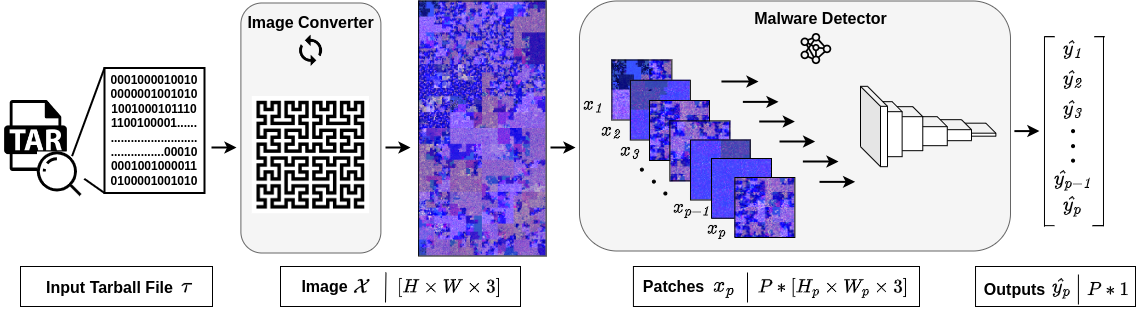}
    \caption{\footnotesize{The proposed method receives a tarball file as input and transforms it into its equivalent RGB image via the Image Converter. Subsequently, the image is split into patches and a convolutional neural network infers whether the input patch representing part of the file system is compromised by malware.}}
    \label{fig:method}
\end{figure*}

\subsection{Deep Learning Malware Detection}
\label{Deep Learning Malware Detection}

The recent advances in Machine Learning have enabled different approaches for detecting malware in different systems. In~\cite{karn2020cryptomining}, the authors propose a system for detecting cryptomining malware within Kubernetes pods in cloud environments. The authors utilize system call monitoring as input to LSTMs~\cite{hochreiter1997long} to identify patterns associated with cryptomining activities. In contrast, our work is broader, targeting diverse malware types within docker containers, enabling attestation in both cloud and local environments. Several studies~\cite{gibert2019using}, and~\cite{cui2018detection} have explored the use of CNNs for malware detection by representing malware binaries as images. There, malware binaries are transformed into grayscale images and classified based on visual features. In contrast, our work addresses the more complex challenge of analyzing complete docker containers rather than individual files, requiring a fundamentally different methodology to capture and classify their behavior. Most similar to our work, Deep-Hook~\cite{landman2021deep} attempts to detect unknown malware that utilizes obfuscation strategies, by dynamically monitoring applications and analyzing volatile memory dumps. Deep-Hook transforms these memory dumps into images for classification using CNNs. Unlike Deep-Hook, which focuses on memory analysis, our approach analyzes the file system, to target malware that exists but remains idle until certain conditions are met, and is thereby not restricted to currently active malware processes.

\section{Proposed Method}

Our aim is to identify dockerized software containers that have been compromised by malware. A common representation of a software container is its associated tarball file which completely captures the system state before execution. As illustrated in Fig.~\ref{fig:method}, the proposed framework initially transforms the tarball file into an equivalent image representation with the Image Converter (described in~\ref{Image Converter}). This step essentially casts the problem of detecting compromised software containers as a binary classification problem on images. Thereafter, the Malware Detector splits the image into patches and classifies them in a streaming, patch-based approach (described in~\ref{Malware Detector}).

\subsection{Image Converter}
\label{Image Converter}
Let each input tarball file $\tau \in \mathcal{T}$ be a collection of files, $\tau=(q_1 \mid q_2 \mid \ldots \mid q_k)$, where each file $ q_i \in \mathcal{Q_\tau}$ is a byte-array of variable length and $(\cdot \mid \cdot)$ is the concatenation operator. Following a common practice for applying machine learning techniques in malware detection, the tarball file $\tau$ is first transformed into a 2D image $x \in \mathcal{X} \hookrightarrow {\rm I\!R}^{h \times w \times c}$, where $h, w, c$ are the height, width, and number of color channels of the image. Converting byte values to pixels is straightforward as both take values in the range $[0-255]$. The layout of the bytes on the image is achieved using space-filling curves. A space-filling curve employs a bijective transformation $Q$ to map a 1-dimensional array to an $n$-dimensional tensor. The transformation can be seen as operating on the indices of the input array. In the case of a 2-dimensional space filling curve, given a $(N \times N)$ square grid with $N^2$ cells, the transformation is a mapping of the form $Q: \{1, ..., N^2\} \longrightarrow N \times N$~\cite{asano_space-filling_1997}. Well-known space-filling curves include zig-zag curves, Hilbert curves, and z-curves, among others. Other image channels can be designed to hold additional information such as byte classification and tarball file structure.

\begin{figure*}[!htbp]
    \centering
    \includegraphics[width=0.99\textwidth]{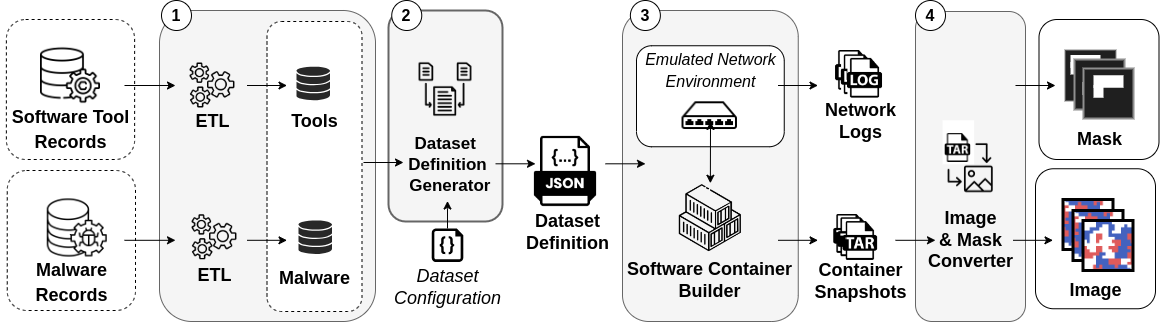}
    \caption{\footnotesize{The data generation pipeline constitutes the following steps: (i) parses tools (1) from the \textit{APT package manager} and malwares from \textit{MalwareBazaar}, (ii) combines them (2) to a dataset definition, (iii) builds each configuration into a container (3) and exports tarball files from the container snapshots along with network traffic logs, and (v) generates filesystem images (4) along with masks highlighting the malware changes.
    }}
    \label{fig:data-generation}
\end{figure*}

\subsection{Malware Detector}
\label{Malware Detector}
The goal is to approximate a function $f: \mathcal{X} \longrightarrow \{0, 1\} $, where $f(x)=1$ indicates that the tarfile contains malware whereas $f(x)=0$ indicates that it does not. This function $f$ is approximated by a convolutional neural network $f_{\theta}$, parameterized by $\theta$, which learns to distinguish between malicious and benign tarball files based on their image representations. The approximated function $f$ needs to be invariant to permutations of the file order inside a tarball, that is if $p(\cdot)$ a valid permutation operating on the tarball underlying file collection $Q_\tau$, then $f(p(\tau)) = f(\tau)$. Although files inside a tarball are usually organized in a hierarchical manner, this structure does not play any significant role in the malware detection task, as a file can be accessed and executed irrespective of its location.

Malware instances operate locally and affect specific bytes within a system whereas the affected bytes are only a small subset compared to the total number of bytes. Therefore, the malware detection problem can be seen under the notion of classification of images based on the information of very small objects or regions of interest (ROIs), in the presence of a much larger and rich background that is uncorrelated with the label~\cite{kong2021efficientclassificationlargeimages}. In addition, the tarball file images are often too large to fit entirely into GPU memory as a whole.

Based on the above, we reformulate the image classification as a streaming, patch-based classification task under the Multiple Instance Learning (MIL)~\cite{Carbonneau_2018} paradigm. In this approach, an image is considered as a bag of patches where if a single patch is identified as containing malicious content, the entire image is classified as malicious. This streaming, patch-based method not only mitigates memory issues associated with larger tarball files but also facilitates early-exit inference. This means that the CNN can terminate the scanning process as soon as the first malicious patch is detected, rather than requiring a full scan of the entire image.  Moreover, our patch-based inference approach provides implicit explainability without reliance on external methods. Upon predicting specific patches as malicious, our approach can directly map these flagged patches back to the exact compromised bytes, providing insights into the location of the malicious content.

Hence, each input tarball file is transformed into a series of image patches $ \{x_p\}_{p=1}^{P} $ where each patch $ x_p \in {\rm I\!R}^{h_p \times w_p \times c}$ corresponds to a portion of the overall image representation of the tarball, and $P$ is the total number of patches. The streaming approach makes predictions on patches sequentially. Specifically, for each patch $x_p$, the CNN outputs $\hat{y_p}=f_{\theta}(x_p)$, where $\hat{y_p} \in \{0, 1\}$ is the predicted label for that patch. The entire tarball is classified as malevolent if at least one patch is predicted as malicious, i.e. $\exists x_p \in x$ s.t. $\hat{y_p}=1$, otherwise the tarball file is classified as benign. Thus, the objective to optimize the $f_{\theta}$ network parameters $\theta$ is the minimization of the binary cross-entropy loss across the individual patches:
\begin{equation*}
    \mathcal{L}(\theta) =  - \frac{ 1 }{ nP  } \sum_{i=1}^{n} \sum_{p=1}^{P}
    [y_{i,p}log(\hat{y}_{i,p}) + (1 - y_{i,p})log(1 - \hat{y}_{i,p})]
\end{equation*}

\section{Dataset Generation Pipeline}

To our knowledge, no public dataset exists that focuses on malware-compromised software containers. To support a machine learning-based detection system, we designed a data generation pipeline that captures the effects of various malware within containerized environments. Each container is initialized with an OS, along with a set of benign packages and tools. Within the container a malware instance is executed in a specialized sandbox environment and the malware is removed. The resulting changes to the file system are recorded using container comparison tools. The final software containers -- both compromised and non-compromised -- are then exported as tarball files and converted to images along with accompanying masks that highlight file-system differences between the two systems on the byte-level.

\subsection{Pipeline Architecture}

The data generation pipeline is designed in a modular fashion to ensure scalability, efficiency and security throughout the data generation process (Fig.~\ref{fig:data-generation}) and consists of the following functionalities organized into components:
\begin{enumerate}
    \item Extract and transform raw data of malware and tools.
    \item Generate dataset definitions from parsed data. 
    \item Build and export docker containers as tarball files. 
    \item Convert tarball files into image - mask pairs.
\end{enumerate}

\subsection{Pipeline Components}


\label{section: Extract-Transform-Load (ETL) malware and tools raw data}
\subsubsection{Extract-Transform-Load (ETL) raw data of Malware and Tools}
The first component ingests raw data on malware records and software packages and tools into the pipeline via ETL processes. This step enables fine-grained control over the selection of malware and tools, taking into account parameters such as targeted CPU architecture, OS, malware families, file types, and characteristics of the benign tools and packages to be installed. The component is also easily extendable to incorporate new malware and software data sources by defining their respective ETL processes.

\label{section: Dataset Definition Generation from Malware and Tools}
\subsubsection{Dataset Definition Generation from Malware and Tools}
This component generates a dataset definition in which each record contains metadata for malware and tool instances, based on the parsed data and user-configured settings. Each record is assigned a unique identifier to support traceback and error handling in case of unexpected behavior during dataset generation. These records are later used by the Software Container Builder to produce container instances, aiming to create a realistic and varied dataset of software containers. Additional constraints can also be applied to the number and characteristics of tool and malware instances per record.


\label{section: Software Container Builder}
\subsubsection{Software Container Builder}
This component receives a dataset definition as input and generates a dataset of software containers as tarball files. It consists of a series of sub-components that rely on the docker Engine to manage the stages of the container-building process: (i) dockerfile generation, (ii) docker image build and (iii) docker container tarball file export. Malware execution within each container occurs in a sandboxed environment with emulated network capabilities. This setup isolates the host system and kernel from malware exposure while simulating network traffic to enable malware activation. After malware activation, the executable is removed from the container. While this may be typical behavior of certain malware variants, it is intentionally enforced in the dataset to constrain the problem scope to detecting the changes inflicted by the malware on the system rather than identifying the malware itself. Moreover, preliminary experiments demonstrated near-perfect accuracy whenn classifying containers that still included the malware executables.
For each malware-compromised container, a corresponding non-compromised container is generated, allowing for byte-level comparisons of malware impact. File changes between compromised and non-compromised containers are tracked using container comparison tools. These differences are stored in a metadata file and used to create masks that highlight the compromised areas.

\label{section: Image & Mask Converter}
\subsubsection{Image \& Mask Converter}
The final component of the pipeline is the converter of tarball files to their image representations. During image creation, other image channels are used to encode supplementary information. Finally, each image representing a compromised software container is paired with a mask of the same dimensions, indicating which files and their respective byte values on the image were affected, i.e., added or modified, by the malware execution.

\subsection{Dataset Generation}

\subsubsection{Data Sources}
Malware Bazaar~\cite{malware_bazaar} is selected as the primary source for malware samples. Malware Bazaar is a popular open-source malware repository that provides a large and diverse collection of malware along with useful metadata. Benign tools and packages are retrieved from the repositories of the Advanced Package Tool (APT) \cite{apt-ubuntu}, which is the basic interface for managing software for Debian-based Linux distributions. Raw data from Malware Bazaar and the APT repositories are ingested through the ETL pipeline component.

\subsubsection{Dataset Statistics}
The dataset statistics are illustrated in Table~\ref{tab:dataset} and Table~\ref{tab:malware}. The dataset consists of 3364 software container records built from 1297 unique packages and 495 unique malware samples spanning 10 malware classes: \textit{Mirai, Gafgyt, CoinMiner, XorDDos, Kaiji, Tsunami, GoBrut, BPFDoor and RotaJakiro} and an extra \textit{Unknown} class with unclassified malware. Each compromised image container is accompanied by a mask indicating the malware compromised bytes. The dataset is split in train, validation and test set in a 70:10:20 split ratio using malware class-based stratified sampling. The exact splits are provided with the dataset.

\vspace{0.2cm}
\begin{table}[!htbp]
    \caption{Generated Dataset Overall Statistics}
    \centering
    \resizebox{0.98\linewidth}{!}{
    \begin{tabular}{l r  r  r  r}
\hline
     & \textbf{Total} & \textbf{Train} & \textbf{Validation} & \textbf{Test}\\
\hline
       Nr. Images              &  3364  &  2360  &   328  &   676 \\
       Nr. Benign Images       &  2225  &  1564  &   214  &   447 \\
       Nr. Compromised Images  &  1139  &   796  &   114  &   229 \\
       Nr. Unique Packages     &  1297  &  1053  &   206  &   393 \\
       Nr. Unique Malware      &   495  &   347  &    49  &   99 \\
       Avg. Image Size         & 158MP  & 158MP  &  157MP & 157MP \\ 

       Avg. Mask / Image Ratio & 0.32\%  & 0.35\%  &  0.29\% & 0.24\% \\ 
       
\hline
    \end{tabular}}
    \label{tab:dataset}
\end{table}

\begin{table}[!htbp]
    \caption{Generated Dataset Malware type Statistics}
    \centering
    \begin{tabular}{l r  r  r}
\hline
     \textbf{Signature} & \textbf{Unique} & \textbf{Total} & \textbf{Avg. Bytes affected}\\
\hline
       Mirai       &  225  &  494  &    58 KB \\
       Gafgyt      &  119  &  284  &   132 KB \\
       CoinMiner   &   28  &   72  &   451 KB \\
       XorDDos     &   27  &   50  &    18 KB \\
       Kaiji       &   21  &   53  &   4.7 MB \\
       Tsunami     &   16  &   43  &   1024 B \\
       GoBrut      &   14  &   34  &    512 B \\
       BPFDoor     &    7  &   16  &    20 KB \\
       RotaJakiro  &    5  &   14  &   134 KB \\
       Unknown     &   33  &   79  &   678 KB \\
\hline
       Total       &  495  & 1139  &   355 KB \\
\hline
    \end{tabular}
    \label{tab:malware}
\end{table}


\subsubsection{Container \& Malware Configuration}

Regarding the targeted system, we opted for an x86x64 CPU architecture, with Ubuntu OS based on the \textit{ubuntu/jammy} as base image. Restrictions on the number of tools of each record were imposed to limit the final tarball file size to be smaller than 200MB. Additional imposed filters included restricting the source APT repositories to exclude documentation-only packages and ubuntu essential packages. As a sandbox container runtime environment, \textit{gVisor}~\cite{gvisor} was selected for its strong isolation capabilities and was executed in Virtual Machines hosted in Google Cloud Platform (GCP).

All produced containers included the \textit{wget} and \textit{p7zip} packages, which were required to facilitate the generation process. The selected malware were limited to \textit{.elf} executables. Malware activation was conducted inside the containers using an emulated network based on \textit{FakeNet-NG 3.2}~\cite{fakenet}. The activation process involved the following steps: (i) Malware executables were downloaded from the Malware Bazaar API and copied into the container. (ii) They were executed and activated as \textit{.elf} binaries and removed afterward. (iii) The resulting software container was compared with its benign counterpart using Google’s \textit{container-diff} tool~\cite{container-diff} and changes on the filesystem - such as additions, modifications and deletions - were recorded. (iv) Software containers that exhibited no file system alteration, or inflicted changes included only deletions were excluded from the dataset.

\subsubsection{Image \& Mask Creation}
The byte-to-pixel value conversion is implemented with an in-house implementation of a byte-to-pixel visualization tool similar to \textit{binvis}~\cite{binvis}. Besides the byte-to-pixel value conversion, additional information was incorporated in other image channels: (a) A byte-class channel, where each pixel gets a color-code based on a crude classification of its value range (i.e., ASCII text, other bytes, and special colors for the \texttt{0x00} and \texttt{0xff} byte values) and the tarball file structure information, where varying colors are assigned to each file in the tarball with additional indicators between the file's header and its content. The pixel layout on the image is performed via a Hilbert space-filling curve due to its clustering properties~\cite{moon_analysis_2001}. Given an image width $w$, the bytes transformed to pixel values are split into $w^2$ chunks which are then mapped to $w \times w$ square patches. This process continues for all bytes. Besides the full image of the tarball file, down-sampled $1024 \times 4096$ and $2048 \times 8192$ images are also provided.

\section{Experiments \& Results}

\subsection{Setup}

To establish a baseline, we begin with an out-of-the-box approach by using VirusTotal~\cite{virus_total}, an online platform that aggregates analyses from over 70 antivirus scanners (including McAfee, Malwarebytes, Kaspersky, etc.) to inspect files and URLs for potential threats. We submit all tarball files of our test set through the VirusTotal API v3 and discuss their results. Then, we implement our method with multiple CNN backbones to evaluate their effectiveness. Specifically, we utilize ResNet18~\cite{he2016deep}, MobileNetV2~\cite{sandler2018mobilenetv2}, and ShuffleNetV2~\cite{ma2018shufflenet}, AlexNet~\cite{krizhevsky2012imagenet} and VGG11~\cite{simonyan2014very}. All models are pretrained on ImageNet~\cite{deng2009imagenet}, which is a standard initialization in computer vision for natural scenes. We adopt this initialization as it has been shown to be particularly effective to textures~\cite{geirhos2018imagenet}, since the RGB byte representation of our dataset produces texture-rich images.


%

Our experiments are performed with PyTorch 1.12 on an Nvidia A6000 for 20 epochs. The Adam optimizer~\cite{kingma2014adam} 
was used with a learning rate of $10^{-4}$ decayed to $10^{-6}$ via the cosine annealing strategy. The networks are trained on the $1024 \times 4096$ version of the images, with a patch size of $256 \times 256$ and a batch size of 256. The loss weights are fixated to 1 and 256 for the negative (benign) and positive (malevolent) classes respectively, to balance the learning gradients between the classes. The images are transformed onto the $[0,1]$ domain to ensure smaller and consistent gradient magnitudes. No geometric augmentations are considered at this stage to avoid distortion of the tarball file representation structure.

\subsection{Evaluation}

Classification performance is evaluated with multiple metrics. 
Accuracy measures the overall correctness of predictions by calculating the proportion of all samples that were classified correctly: 
\begin{equation*}
Accuracy = \frac{TP + FP}{TP+TN+FN+FP}
   \nonumber
\end{equation*}

While accuracy gives an overall sense of performance, it may be misleading when different errors has different costs. In our case, misclassifying a malevolent container as benign introduces more problems than misclassifying benign containers, as it could breach the entire security layer. For this reason, our study includes the accuracy metric, yet focuses on precision, recall, and f1 score, specifically for the positive (malevolent) class. Precision for the positive class calculates the proportion of correctly predicted positive samples out of all samples predicted as positive, whereas recall, also known as sensitivity or true positive rate, measures the proportion of actual positive samples that were correctly identified. The F1 score combines precision and recall into a single metric, i.e. it is the ``harmonic mean" of precision and recall.

\begin{equation*}
Precision = \frac{TP}{TP+FP}, \qquad Recall = \frac{TP}{TP+FN}
    \nonumber
\end{equation*}


\begin{equation*}
F1\ score = 2 \times \frac{Precision \times Recall}{Precision + Recall}
    \nonumber
\end{equation*}

\begin{table}[!htbp]
    \caption{Test set performance of various methods.}
    \centering
    \begin{tabular}
    {p{2.1cm} |c  c  c  c  | p{0.5cm} }
\hline
       \textbf{Method}    & 
       \textbf{F1}        &  
       \textbf{Precision} &  
       \textbf{Recall}    & 
       \textbf{Accuracy}  & 
       \textbf{\#P(M)} \\
\hline
       Panda (\#32)       & 0.085 & 1.000 & 0.044 & 0.679 & N/A \\
       Kaspersky (\#18)   & 0.598 & 1.000 & 0.427 & 0.807 & N/A \\
       BitDefender (\#10) & 0.624 & 1.000 & 0.453 & 0.816 & N/A \\
       Ikarus (\#1)       & 0.703 & 1.000 & 0.542 & 0.846 & N/A \\
       VirusTotal ($\geq20$) & 0.601 & 0.990 & 0.431 & 0.807 & N/A \\       
       VirusTotal ($\geq1$) & 0.709 & 0.992 & 0.551 & 0.848 & N/A \\
\hline
       VGG11            & 0.552 & 0.928 & 0.393 & 0.784  & 132.9\\
       ShuffleNetV2     & 0.586 & 0.857 & 0.445 & 0.787  & 2.3\\
       MobileNetV2      & 0.622 & 0.855 & 0.489 & 0.799  & 3.5\\
       AlexNet          & 0.680 & 0.821 & 0.581 & 0.815  & 61.1\\
       EfficientNet     & 0.724 & 0.861 & 0.624 & 0.837  & 5.3\\
       ResNet18         & 0.736 & 0.826 & 0.664 & 0.839  & 11.7\\
\hline
    \end{tabular}
    \label{tab:results}
\end{table}

\subsection{Results}

The experimental findings are illustrated in Tab.~\ref{tab:results}. Interestingly, all scanners have near perfect precision, yet they suffer from false negatives as indicated by low recall scores. Ikarus achieves the highest recall with an F1 score of 0.703. We further demonstrate the rankings for other well-known engines such as BitDefender and Kaspersky. Last we present metrics for ``ensembles of engines" where we aggregate their predictions based on simple voting strategies. We denote with $\geq a$ the results where at least $a$ engines classify a record as malevolent. The complete analysis across all engines and ensembles will be provided along with the dataset. 


The results of our method with ResNet18 indicate that it stands out in terms of performance with an F1 score of 0.736. Our method outperforms all individual engines as well as all their ensembles while being a lot faster than the API call and not relying on external software. EfficientNet performs slightly worse, i.e. achieves an F1 score of 0.726, slightly sacrificing Recall for higher Precision, yet requires less than only half of the parameters. Interestingly, AlexNet follows with an F1 score of 0.680, despite the relatively older architectural design. MobileNetV2 and ShuffleNetV2 perform slightly worse but they use orders of magnitude less parameters than VGG11 that achieves the lowest F1 score among the CNNs. 

\begin{figure}[!bp]
\centering
\begin{subfigure}{0.15\linewidth}
    \centering
    \includegraphics[width=\textwidth]{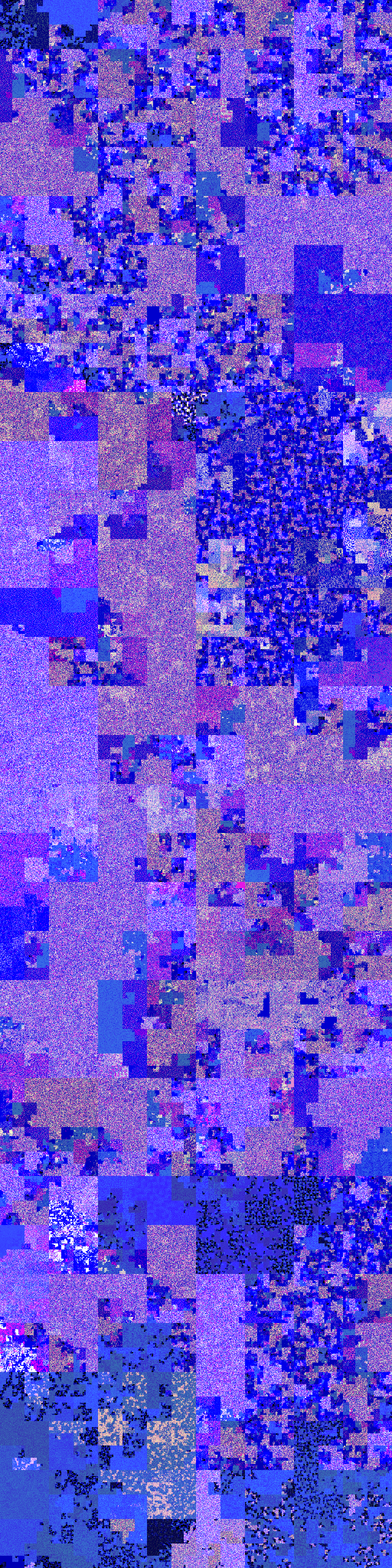}
\end{subfigure}%
\hfill
\begin{subfigure}{0.15\linewidth}
    \centering
    \includegraphics[width=\textwidth]{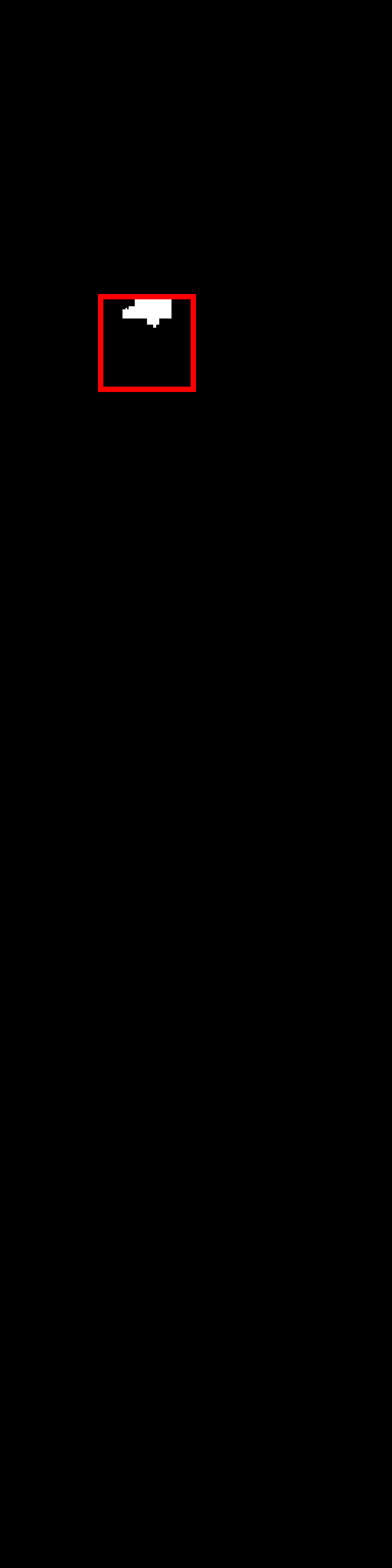}
\end{subfigure}%
\hfill
\begin{subfigure}{0.15\linewidth}
    \centering
    \includegraphics[width=\textwidth]{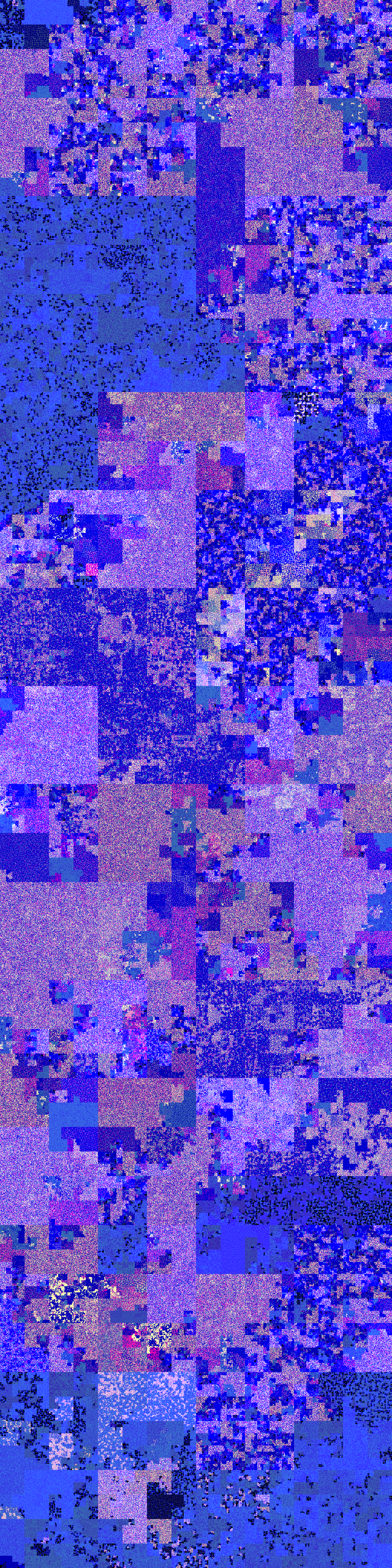}
\end{subfigure}%
\hfill
\begin{subfigure}{0.15\linewidth}
    \centering
    \includegraphics[width=\textwidth]{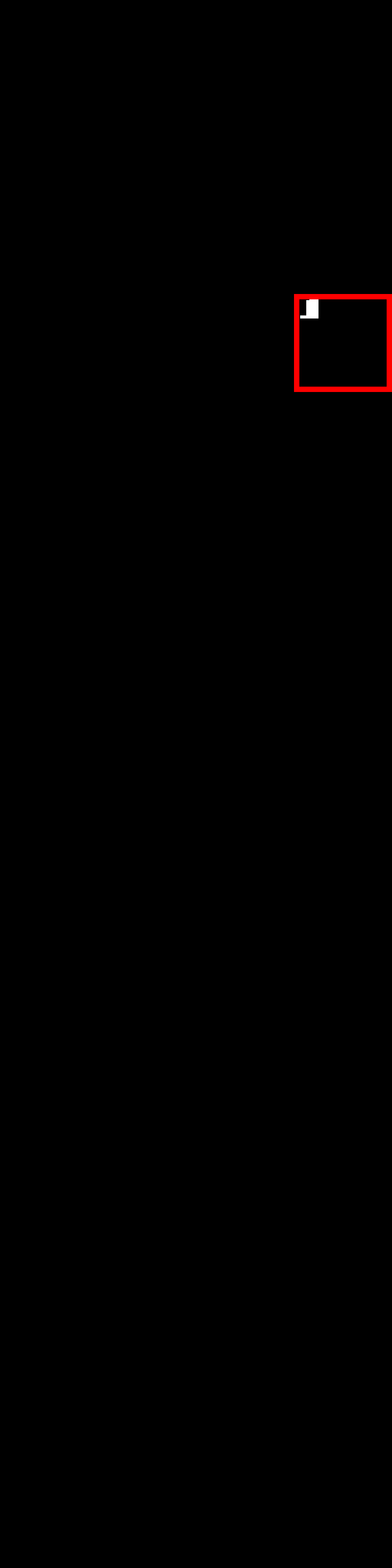}
\end{subfigure}%
\hfill
\begin{subfigure}{0.15\linewidth}
    \centering
    \includegraphics[width=\textwidth]{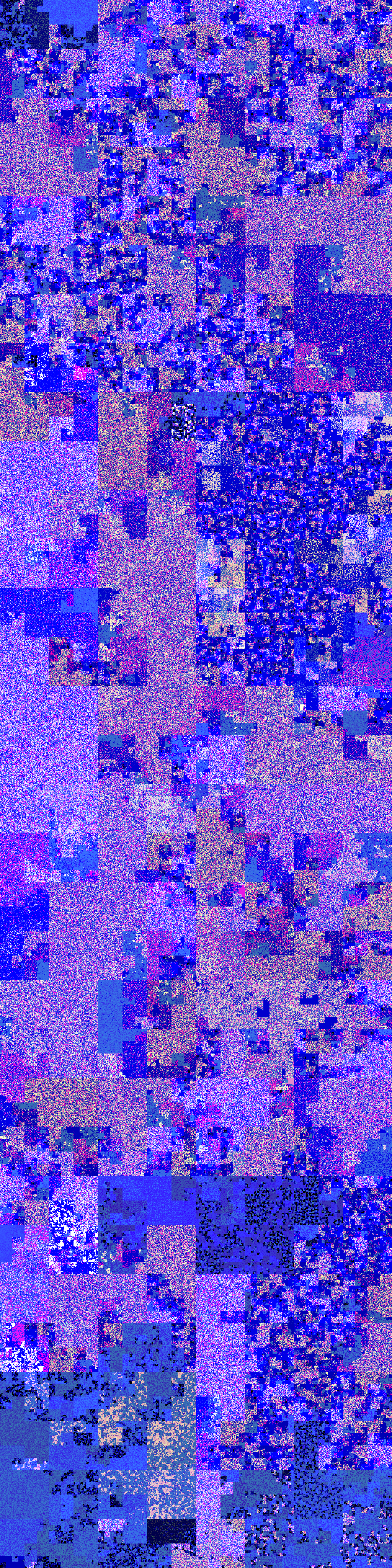}
\end{subfigure}%
\hfill
\begin{subfigure}{0.15\linewidth}
    \centering
    \includegraphics[width=\textwidth]{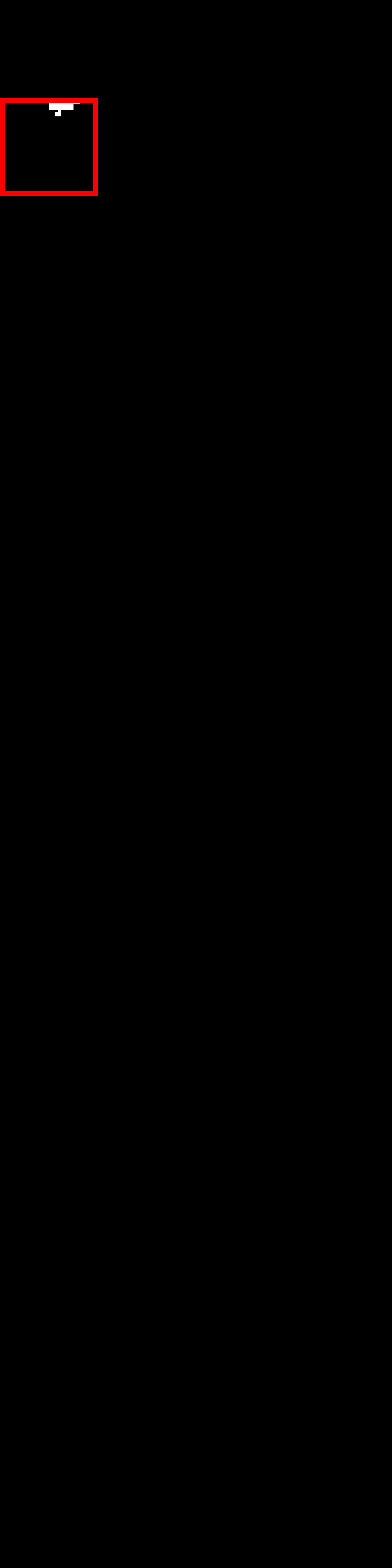}
\end{subfigure}%

\caption{\footnotesize{Qualitative illustration. Each pair illustrates an image and its respective mask. White pixels refer to the bytes compromised by malware activity, whereas the red bounding boxes indicate the patch predicted as malevolent using ResNet18. Not only does our method correctly classify images, but does so by detecting the patch of interest. Our patch-based approach a) explicitly reduces GPU memory required and b) implicitly allows for i) early-exit inference to improve the runtime and ii) explanations for which patches contributed to the prediction.}}
\label{fig:qualitative}
\end{figure}

Additionally, as our patch-based framework dissects the malware detection process by extracting decisions for each input patch, it naturally provides insights into not only the overall malevolence of the image but also the specific patches contributing to it. As illustrated in Fig.~\ref{fig:qualitative}, our framework effectively identifies the patch with malicious content, providing a notion of explainability. By reverse engineering the image generation process using the Hilbert space-filling permutation, we can derive the exact byte locations of the malicious component within the container.


\section{Discussion}

Overall, our results underscore that end-to-end malware detection in docker containers is challenging. Both \textit{F1 score} and \textit{recall} metrics highlight the difficulty of detecting the patterns introduced by malware across large file systems. Nonetheless, the effective application of deep learning architectures shows promise and opens up new avenues for end-to-end security systems that consider the underlying file system and leverage the generalization capabilities of deep neural networks.

Despite the successful experiment outcomes, more work is required to study the workings of the proposed method. For instance, our work transforms the tarball files into images, following common prior practices. However, there is no evidence on how this representation is better than some other. Recent work~\cite{wu2024beyond}, ~\cite{pagnoni2024byte} has showed that transformers can be directly trained from byte-level input representations opening interesting research avenues for foundation models on byte-level data. Furthermore, our method is trained on the down-sampled, $1024\times 4096$ input images. While this might be more optimal in terms of runtime efficiency, it might be compromising malware detection performance due to the aggregation of the input signal. Naturally, future work will examine the performance versus efficiency trade-offs at higher resolutions and investigate techniques under the framework of Multiple Instance Learning that may be more suitable. Moreover, future work involves extending the task of identifying compromised software containers as a pixel-level semantic segmentation task, which among others can provide direct insights into the affected bytes.

\section{Conclusion}

In this paper, our efforts were focused in further supporting the transition of malware detection from traditional methods to machine learning-based techniques as well as tackle the need of safeguarding container-based systems. We introduced the task of identifying compromised software containers given their file system with machine learning and proposed a streaming, patch-based CNN approach. To support our work and encourage further research in the field, we release the Compromised Software Containers (COSOCO) image dataset, with benign and compromised docker containers. Finally, we used the COSOCO dataset to train CNN classification models and produce baseline results.

\bibliographystyle{unsrt}
\bibliography{bibl}

\end{document}